\documentstyle[twoside,fleqn,espcrc2]{article}

\def\beq{\begin{equation}} 
\def\eeq{\end{equation}} 
\def\bea{\begin{eqnarray}} 
\def\eea{\end{eqnarray}}
\def\bq{\begin{quote}} 
\def\eq{\end{quote}}

\def\gappeq{\mathrel{\rlap {\raise.5ex\hbox{$>$}} {\lower.5ex\hbox{$\sim$}}}}

\def\lappeq{\mathrel{\rlap{\raise.5ex\hbox{$<$}} {\lower.5ex\hbox{$\sim$}}}}

\newcommand{\AmS}{{\protect\the\textfont2
  A\kern-.1667em\lower.5ex\hbox{M}\kern-.125emS}}

\title{THE STANDARD MODEL AND BEYOND}

\author{G. Altarelli\address{Theoretical Physics Division, CERN \\ 
        1211 Geneva 23, Switzerland\\
        and\\
        Universit\`a di Roma Tre, Rome, Italy}
  }     
\begin{document}

\begin{abstract}
\noindent
{\bf Content}\\
1.Why we do Believe in the Standard Model\\
2.Why we do not Believe in the Standard Model\\
\indent 
2.1.Conceptual Problems \\
\indent 
2.2.Hints from Experiment \\
\indent --2.2.1 Unification of Couplings\\
\indent 
--2.2.2 Dark Matter\\
\indent --2.2.3
Baryogenesis\\
\indent --2.2.4 Neutrino Masses
\\
\indent --2.2.5 Ultra High Energy Cosmic Rays
\\3.Conclusion
\end{abstract}
\maketitle



\section{Why we do Believe in the SM: Precision Tests}

In recent years new powerful tests of the Standard Model (SM) have been performed mainly at LEP but also
at SLC and at the Tevatron. The running of LEP1 was terminated in 1995 and close-to-final results of the data
analysis are now available \cite{tim},\cite{ew}. The experiments at the Z resonance have enormously improved
the accuracy of the data in the electroweak neutral current sector \cite{sta}. The top quark has been at last
found and the errors on $m_Z$ and $\sin^2\theta_{eff}$ went down by two and one orders of magnitude
respectively since the start of LEP in 1989. The LEP2 programme is in progress. The validity
of the SM has been confirmed to a level that we can say was unexpected at the beginning. In the present data
there is no significant evidence for departures from the SM, no convincing hint of new physics (also
including the sofar available results from LEP2) \cite{tre}. The impressive success of the SM poses strong
limitations on the possible forms of new physics. Favoured are models of the Higgs sector and of new physics that
preserve the SM structure  and only very delicately improve it, as is the case for fundamental Higgs(es) and
Supersymmetry. Disfavoured are models with a nearby strong non perturbative regime that  almost inevitably
would affect the radiative corrections, as for composite Higgs(es) or technicolor and its variants. 

The main results of the precision tests of the standard electroweak theory can be summarised as follows. It has
been checked that the couplings of quark and leptons to the weak gauge bosons $W^{\pm}$ and $Z$ are indeed
precisely as prescribed by the gauge symmetry. The accuracy of a few 0.1\% for these tests implies that, not
only the tree level, but also the structure of quantum corrections has been verified. To a lesser accuracy the
triple gauge vertices
$\gamma W^+ W^-$ and
$Z W^+ W^-$ have also been found in agreement with the specific prediction of the $SU(2)\bigotimes U(1)$ gauge
theory, at the tree level. This means that we have verified that the gauge symmetry is indeed unbroken in the
vertices of the theory: the currents are indeed conserved. Yet we have immediate evidence that the symmetry is
otherwise badly broken in the masses. In fact the $SU(2)\bigotimes U(1)$ gauge symmetry forbids masses for all
the particles that have been sofar observed: quarks, leptons and gauge bosons. Of all these particles
only the photon is massless (and the gluons protected by the $SU(3)$ colour gauge symmetry), all other are massive
(probably also the neutrinos). Thus the currents are conserved but the particle states are not symmetric. This is
the definition of spontaneous symmetry breaking. The simplest implementation of spontaneous symmetry breaking in a
gauge theory is via the Higgs mechanism. In the Minimal Standard Model (MSM) one single scalar Higgs isospin
doublet is introduced and its vacuum expectation value v breaks the symmetry. All masses are proportional to v,
although for quarks and leptons the spread of the Yukawa couplings that multiply v in the expression for the masses
are distributed over a wide range. The Higgs sector is still largely untested. The Higgs particle has not been
found: being coupled in proportion to masses one has first to produce heavy particles and then try to detect the
Higgs (itself heavy) in their couplings. The present limit is $m_H\gappeq m_Z$ from LEP. What has been tested is
the relation $m^2_W=m^2_Z \cos^2{\theta_W}$, modified by computable radiative corrections. This relation means that
the effective Higgs (be it fundamental or composite) is indeed a weak isospin doublet. Quantum corrections to the
electroweak precision tests depend on the masses and the couplings in the theory. For example they depend on the top
mass
$m_t$, the Higgs mass
$m_H$, the strong coupling $\alpha_s(m_Z)$, the QED coupling $\alpha(m_Z)$ (these are running couplings at the Z mass)
and other parameters which are better known. In particular quantum corrections depend quadratically on $m_t$ and only
logaritmically on $m_H$. From the observed radiative corrections one obtains a value of $m_t$ in fair agreement
with the observed value from the TeVatron. For the Higgs mass one finds $\log_{10}{m_H(GeV)}=1.92^{+0.32}_{-0.41}$
(or $m_H=84^{+91}_{-51}$). This result on the Higgs mass is particularly remarkable. Not only the value of
$\log_{10}{m_H(GeV)}$ is right on top of the small window between $\sim 2$ and $\sim 3$ which is allowed by the
direct limit, on the one side, and the theoretical upper limit on the Higgs mass in the MSM,
$m_H\lappeq 800~GeV$, on the other side. If one had found a central value like $\gappeq 4$ the model would have
been discarded. Thus the whole picture of a perturbative theory with a fundamental Higgs is well supported by the
data. But also there is clear indication for a particularly light Higgs. This is quite encouraging for the ongoing
search for the Higgs particle. More in general, if the Higgs couplings are removed from the lagrangian the
resulting theory is non renormalisable. A cutoff $\Lambda$ must be introduced. In the quantum corrections 
$\log{m_H}$ is then replaced by $\log{\Lambda}$. The precise determination of the associated finite terms would be
lost (that is, the value of the mass in the denominator in the argument of the logarithm). But the generic
conclusion would remain, that, whatever the mechanism of symmetry breaking, the experimental solution of the
corresponding problem, is not far away in energy.

\section{Why we do not Believe in the SM}
\subsection{Conceptual Problems}

	Given the striking success of the SM why are we not satisfied with that theory? Why not just find the Higgs
particle, for completeness, and declare that particle physics is closed? The main reason is that there are
strong conceptual indications for physics beyond the SM. 

	It is considered highly unplausible that the origin of the electro-weak symmetry breaking can be explained by
the standard Higgs mechanism, without accompanying new phenomena. New physics should be manifest at energies in
the TeV domain. This conclusion follows fron an extrapolation of the SM at very high energies. The computed
behaviour of the $SU(3)\otimes SU(2)\otimes U(1)$ couplings with energy clearly points towards the
unification of the electro-weak and strong forces (Grand Unified Theories:
GUTs) at scales of energy
$M_{GUT}\sim  10^{14}-10^{16}~ GeV$ which are close to the scale of quantum gravity, $M_{Pl}\sim 10^{19}~ GeV$
\cite{qqi}.  One can also imagine  a unified theory of all interactions also including gravity (at
present superstrings
\cite{ler} provide the best attempt at such a theory). Thus GUTs and the realm of quantum gravity set a
very distant energy horizon that modern particle theory cannot anymore ignore. Can the SM without new physics be
valid up to such large energies? This appears unlikely because the structure of the SM could not naturally
explain the relative smallness of the weak scale of mass, set by the Higgs mechanism at $\mu\sim
1/\sqrt{G_F}\sim  250~ GeV$  with $G_F$ being the Fermi coupling constant. This so-called hierarchy problem
\cite{ssi} is related to the presence of fundamental scalar fields in the theory with quadratic mass divergences
and no protective extra symmetry at $\mu=0$. For fermions, first, the divergences are logaritmic and, second, at
$\mu=0$ an additional symmetry, i.e. chiral  symmetry, is restored. Here, when talking of divergences we are not
worried of actual infinities. The theory is renormalisable and finite once the dependence on the cut off is
absorbed in a redefinition of masses and couplings. Rather the hierarchy problem is one of naturalness. If we
consider the cut off as a manifestation of new physics that will modify the theory at large energy scales, then it
is relevant to look at the dependence of physical quantities on the cut off and to demand that no unexplained
enormously accurate cancellations arise. 

	According to the above argument the observed value of $\mu\sim 250~ GeV$ is indicative of the existence of new
physics nearby. There are two main possibilities. Either there exist fundamental scalar Higgses but the theory
is stabilised by supersymmetry, the boson-fermion symmetry that would downgrade the degree of divergence from
quadratic to logarithmic. For approximate supersymmetry the cut off is replaced by the splitting between the
normal particles and their supersymmetric partners. Then naturalness demands that this splitting (times the
size of the weak gauge coupling) is of the order of the weak scale of mass, i.e. the separation within
supermultiplets should be of the order of no more than a few TeV. In this case the masses of most supersymmetric
partners of the known particles, a very large managerie of states, would fall, at least in part, in the discovery
reach of the LHC. There are consistent, fully formulated field theories constructed on the basis of this idea, the
simplest one being the Minimal Supersymmetric Standard Model (MSSM) \cite{43}. As already mentioned, all normal observed states are those whose masses are
forbidden in the limit of exact
$SU(2)\otimes U(1)$. Instead for all SUSY partners the masses are allowed in that limit. Thus when
supersymmetry is broken in the TeV range but $SU(2)\otimes U(1)$ is intact only s-partners take mass while all
normal particles remain massless. Only at the lower weak scale the masses of ordinary particles are generated.
Thus a simple criterium exists to understand the difference between particles and s-particles.

	The other main avenue is compositeness of some sort. The Higgs boson is not elementary but either a bound
state of fermions or a condensate, due to a new strong force, much stronger than the usual strong interactions,
responsible for the attraction. A plethora of new "hadrons", bound by the new strong force would  exist in the
LHC range. A serious problem for this idea is that nobody sofar has been  able to build up a realistic model
along these lines, but that could eventually be explained by a lack of ingenuity on the theorists side. The
most appealing examples are technicolor theories \cite{30},\cite{chi}. These models were inspired by the
breaking of chiral symmetry in massless QCD induced by quark condensates. In the case of the electroweak
breaking new heavy techniquarks must be introduced and the scale analogous to $\Lambda_{QCD}$ must be about
three orders of magnitude larger. The presence of such a large force relatively nearby has a strong tendency to
clash with the results of the electroweak precision tests \cite{32}.

	The hierarchy problem is certainly not the only conceptual problem of the SM. There are many more: the
proliferation of parameters, the mysterious pattern of fermion masses and so on. But while most of these
problems can be postponed to the final theory that will take over at very large energies, of order $M_{GUT}$ or
$M_{Pl}$, the hierarchy problem arises from the unstability of the low energy theory and requires a solution at
relatively low energies. 

A supersymmetric extension of the SM provides a way out which is well defined,
computable and that preserves all virtues of the SM.  The necessary SUSY breaking can be introduced through soft
terms that do not spoil the good convergence properties of the theory. Precisely those terms arise from
supergravity when it is spontaneoulsly broken in a hidden sector \cite{yyi}. But alternative mechanisms of SUSY
breaking are also being considered
\cite{gauge}.  In the
most familiar approach SUSY is broken in a hidden sector and the scale of SUSY breaking is very
large of order
$\Lambda\sim\sqrt{G^{-1/2}_F M_P}$  where
$M_P$ is the Planck mass. But since the hidden sector only communicates with the visible sector
through gravitational interactions the splitting of the SUSY multiplets is much smaller, in the TeV
energy domain, and the Goldstino is practically decoupled. In an alternative scenario the (not so
much) hidden sector is connected to the visible one by ordinary gauge interactions. As these are much
stronger than the gravitational interactions, $\Lambda$ can be much smaller, as low as 10-100
TeV. It follows that the Goldstino is very light in these models (with mass of order or below 1 eV
typically) and is the lightest, stable SUSY particle, but its couplings are observably large. The radiative
decay of the lightest neutralino into the Goldstino leads to detectable photons. The signature of photons comes
out naturally in this SUSY breaking pattern: with respect to the MSSM, in the gauge mediated model there are typically
more photons and less missing energy. Gravitational and gauge mediation are extreme alternatives: a spectrum
of intermediate cases is conceivable. The main appeal of gauge mediated models is a better protection against
flavour changing neutral currents. In the gravitational version even if we accept that gravity leads to
degenerate scalar masses at a scale near $M_{Pl}$ the running of the masses down to the weak scale can
generate mixing induced by the large masses of the third generation fermions \cite{ane}.

\subsection{Hints from Experiment}
\subsubsection{Unification of Couplings}

At present the most direct
phenomenological evidence in favour of supersymmetry is obtained from the unification of couplings in
GUTs.
Precise LEP data on $\alpha_s(m_Z)$ and $\sin^2{\theta_W}$ confirm what was already known with less accuracy:
standard one-scale GUTs fail in predicting $\sin^2{\theta_W}$ given
$\alpha_s(m_Z)$ (and $\alpha(m_Z)$) while SUSY GUTs \cite{zzi} are in agreement with the present, very precise,
experimental results. According to the recent analysis of ref\cite{aaii}, if one starts from the known values of
$\sin^2{\theta_W}$ and $\alpha(m_Z)$, one finds for $\alpha_s(m_Z)$ the results:
\bea
\alpha_s(m_Z) &=& 0.073\pm 0.002 ~(\rm{Standard~ GUTs})\nonumber \\	
		\alpha_s(m_Z) &=& 0.129\pm0.010~(\rm{SUSY~ GUTs})
\label{24}
\eea
to be compared with the world average experimental value $\alpha_s(m_Z)$ =0.119(4).

\subsubsection{Dark Matter}

There is solid astrophysical and cosmological evidence \cite{kol}, \cite{spi} that most of the matter in the universe
does not emit electromagnetic radiation, hence is "dark". Some of the dark matter must be baryonic but most of it must
be non baryonic. Non baryonic dark matter can be cold or hot. Cold means non relativistic at freeze out, while hot is
relativistic. There is general consensus that most of the non baryonic dark matter must be cold dark matter. A couple
of years ago the most likely composition was quoted to be around 80\% cold and
20\% hot. At present it appears to me
that the need of a sizeable hot dark matter component is more uncertain. In fact, recent experiments have indicated the
presence of a previously disfavoured cosmological constant component in
$\Omega=\Omega_m+\Omega_{\Lambda}$ \cite{kol}. Here
$\Omega$ is the total matter-energy density in units of the critical density, $\Omega_m$ is the matter component
(dominated by cold dark matter) and $\Omega_{\Lambda}$ is the cosmological component. Inflationary theories almost
inevitably predict
$\Omega=1$ which is consistent with present data. At present, still within large uncertainties, the approximate
composition is indicated to be
$\Omega_m\sim 0.4$ and
$\Omega_{\Lambda}\sim0.6$ (baryonic dark matter gives $\Omega_b\sim0.05$). 

The implications for particle physics is that certainly there must exist a source of cold dark matter. By far the
most appealing candidate is the neutralino, the lowest supersymmetric particle, in general a superposition of
photino, Z-ino and higgsinos. This is stable in supersymmetric models with R parity conservation, which are the
most standard variety for this class of models (including the MSSM). A
neutralino with mass of order 100 GeV would fit perfectly as a cold dark matter candidate. Another common
candidate for cold dark matter is the axion, the elusive particle associated to a possible solution of the strong
CP problem along the line of a spontaneously broken Peccei-Quinn symmetry. To my knowledge and taste this option is
less plausible than the neutralino. One favours supersymmetry for very diverse conceptual and
phenomenological reasons, as described in the previous sections, so that neutralinos are sort of standard by now.
For hot dark matter, the self imposing candidates are neutrinos. If we demand a density fraction
$\Omega_{\nu}\sim0.1$ from neutrinos, then it turns out that the sum of stable neutrino masses should be around 5
eV. 

\subsubsection{Baryogenesis}

 Baryogenesis is interesting because it could occur at the weak
scale \cite{rub} but not in the SM. For baryogenesis one needs the three famous Sakharov conditions \cite{sak}: B
violation, CP violation and no termal equilibrium. In principle these conditions could be verified in the SM. B is
violated by instantons when kT is of the order of the weak scale (but B-L is conserved). CP is violated by the CKM
phase and out of equilibrium conditions could be verified during the electroweak phase transition. So the
conditions for baryogenesis appear superficially to be present for it to occur at the weak scale in the SM.
However, a more quantitative analysis \cite{rev}, \cite{cw1} shows that baryogenesis is not possible
in the SM because there is not enough CP violation and the phase transition is not sufficiently strong first order,
unless
$m_H<80~GeV$, which is by now excluded by LEP. Certainly baryogenesis could also occur  below the
GUTs scale, after
inflation. But only that part with
$|B-L|>0$ would survive and not be erased at the weak scale by instanton effects. Thus baryogenesis at $kT\sim
10^{12}-10^{15}~GeV$ needs B-L violation at some stage like for $m_\nu$. The two effects could be related if
baryogenesis arises from leptogenesis \cite{lg} then converted into baryogenesis by instantons. While baryogenesis
at a large energy scale is thus not excluded it is interesting that recent studies have shown that baryogenesis at
the weak scale could be possible in the MSSM \cite{cw1}. In fact, in this model there are additional sources of CP
violations and the bound on $m_H$ is modified by a sufficient amount by the presence of scalars with large
couplings to the Higgs sector, typically the s-top. What is required is that
$m_h\sim 80-110~GeV$ (in the
LEP2 range!), a s-top not heavier than the top quark and, preferentially, a small $\tan{\beta}$.

\subsubsection{Neutrino Masses}

Recent data from Superkamiokande \cite{SK}(and also MACRO \cite{MA}) have provided a more
solid experimental basis for neutrino oscillations as an explanation of the atmospheric neutrino
anomaly. In addition the solar neutrino deficit is also probably an indication of a different
sort of neutrino oscillations. Results from the laboratory experiment by the LNSD
collaboration \cite{LNSD} can also be considered as a possible indication of yet another type
of neutrino oscillation. But the preliminary data from Karmen \cite{KA} have failed to
reproduce this evidence. The case of LNSD oscillations is far from closed but one can
tentatively assume, pending the results of continuing experiments, that the signal will not
persist. Then solar and atmospheric neutrino oscillations can possibly be explained in terms
of the three known flavours of neutrinos without invoking extra sterile species. Neutrino
oscillations for atmospheric neutrinos require
$\nu_{\mu}\rightarrow\nu_{\tau}$ with $\Delta m^2_{atm}\sim 2~10^{-3}~eV^2$ and a nearly
maximal mixing angle
$\sin^2{2\theta_{atm}}\geq 0.8$. In most of the Superkamiokande allowed region the bound by Chooz
\cite{Chooz} essentially excludes $\nu_e\rightarrow\nu_{\mu}$ oscillations for atmospheric neutrino
oscillations. Furthermore the last results from Superkamiokande allow a solution of the
solar neutrino deficit in terms of
$\nu_e$ disappearance vacuum oscillations (as opposed to MSW \cite{MSW} oscillations within the sun)
with $\Delta m^2_{sol}\sim ~10^{-10}~eV^2$ and again nearly maximal mixing angles. Among the
large and small angle MSW solutions the small angle one is perhaps more likely  at the moment
(with \cite{Bahcall} $\Delta m^2_{sol}\sim 0.5~10^{-5}~eV^2$ and $\sin^2{2\theta_{sol}}\sim
5.5~10^{-3}$) than the large angle MSW solution. Of course experimental uncertainties are
still large and the numbers given here are merely indicative. But by now it is very unlikely that all this
evidence for neutrino oscillations will disappear or be explained away by astrophysics or other solutions. The
consequence is that we have a substantial evidence that neutrinos are massive.

In a strict minimal standard model point of view neutrino masses could vanish if no right handed neutrinos
existed (no Dirac mass) and lepton number was conserved (no Majorana mass). In
GUTs both these
assumptions are violated. The right handed neutrino is required in all unifying groups larger than SU(5). In SO(10)
the 16 fermion fields in each family, including the right handed neutrino, exactly fit into the 16 dimensional
representation of this group. This is really telling us that there is something in SO(10)! The SU(5)
alternative in terms of $\bar 5+10$, without a right handed neutrino, is certainly less elegant. The breaking of
$|B-L|$, B and L is also a generic feature of GUTs. In fact, the see-saw mechanism \cite{ssm} explains
the smallness of neutrino masses in terms of the large mass scale where $|B-L|$ and L are violated. Thus, neutrino
masses, as would be proton decay, are important as a probe into the physics at the
GUTs scale.

Oscillations only determine squared mass differences and not masses. The case of three nearly degenerate neutrinos
is the only one that could in principle accomodate neutrinos as hot dark matter together with solar and atmospheric
neutrino oscillations. According to our previous discussion, the common mass should be around 1-3 eV. The solar
frequency could be given by a small 1-2 splitting, while the atmospheric frequency could be given by a still small
but much larger 1,2-3 splitting. A strong constraint arises in the degenerate case from neutrinoless double beta
decay which requires that the ee entry of
$m_{\nu}$ must obey
$|(m_{\nu})_{11}|\leq 0.46~{\rm eV}$. As observed in ref. \cite{GG}, this bound can only be 
satisfied if
double maximal mixing is realized, i.e. if also solar neutrino oscillations occur with nearly maximal mixing.
We have mentioned that it is not at all clear at the moment that a hot dark matter component is really
needed \cite{kol}. However the only reason to consider the fully degenerate solution is 
that it is compatible
with hot dark matter.
Note that for degenerate masses with $m\sim 1-3~{\rm eV}$ we need a relative splitting $\Delta m/m\sim
\Delta m^2_{atm}/2m^2\sim 10^{-3}-10^{-4}$ and an even smaller one for solar neutrinos. We were unable
to imagine a natural mechanism compatible with unification and the see-saw mechanism to arrange such a
precise near symmetry.

If neutrino masses are smaller than for cosmological relevance, we can have the hierarchies $|m_3| >> |m_{2,1}|$
or $|m_1|\sim |m_2| >> |m_3|$. Note that we
are assuming only two frequencies, given by $\Delta_{sun}\propto m^2_2-m^2_1$ and
$\Delta_{atm}\propto m^2_3-m^2_{1,2}$. We prefer the first case, because for quarks and leptons one
mass eigenvalue, the third generation one, is largely dominant. Thus the dominance of $m_3$ for neutrinos
corresponds to what we observe for the other fermions.  In this case, $m_3$ is determined by the atmospheric
neutrino oscillation frequency to be around $m_3\sim0.05~eV$. By the see-saw mechanism $m_3$ is related to some
large mass M, by $m_3\sim m^2/M$. If we identify m with either the Higgs vacuum expectation value or the top mass
(which are of the same order), as suggested for third generation neutrinos by
GUTs in simple SO(10)
models, then M turns out to be around $M\sim 10^{15}~GeV$, which is consistent with the connection with
GUTs. If
solar neutrino oscillations are determined by vacuum oscillations, then $m_2\sim 10^{-5}~eV$ and we have that the
ratio $m_2/m_3$ is well consistent with $(m_c/m_t)^2$.

A lot of attention \cite{gaff} is being devoted to the
problem of a natural explanation of the observed nearly maximal mixing angle for atmospheric
neutrino oscillations and possibly also for solar neutrino oscillations, if explained by vacuum
oscillations. Large mixing angles are somewhat unexpected because
the observed quark mixings are small and the quark, charged lepton and neutrino mass matrices are to
some extent related in GUTs. There must be some special interplay between the neutrino Dirac
and Majorana matrices in the see-saw mechanism in order to generate maximal
mixing. It is hoped that looking for a natural explanation of large neutrino mixings can lead us to decripting
some interesting message on the physics at the GUT scale.

\subsubsection{Ultra High Energy Cosmic Rays}

The observation by the Fly's Eye and AGASA collaborations of proton-like cosmic rays with energies of order $\gappeq
10^{11}~GeV$ well above the GKZ cutoff of a few $10^{10}~GeV$ poses serious problems in terms of a possible
astrophysical explanation. The GKZ cutoff arises from absorption of protons by the cosmic microwave background if the
proton energy is sufficient to induce N* resonant photoproduction of pions. For these energetic protons the mean free
path in space is limited to the vicinity of our galaxy, say to a distance of order of 50Mpc. On the other hand the
angular distribution of high energy proton events indicates their extragalactic origin. So the problem is either to find
sufficiently energetic nearby astrophysical sources or to explain the observed events in terms of some new effect in
particle physics. I understand that an astrophysical solution is still not excluded
($\gamma$-ray bursts?). As far as a possible particle
physics explanation is concerned one class of solutions is based on assuming the UHECR are not protons, but some
exotic hadron-like particle. The least unplausible example of such a particle is a hadron with light gluino
constituents \cite{far}. This hadron of larger mass than the proton would probably have a smaller crossection for pion
photoproduction and evade the GFK bound. Other solutions like a neutrino with enormously enhanced crosssections at
small x are untenable \cite{hal}. One different possibility is if primary cosmic neutrinos annihilate with cosmic
background neutrinos to produce a Z which then decays into protons \cite{wei}. A relic neutrino of mass of order few eV
would be needed. But the problems are that one requires a very large flux of neutrinos of very high energy
\cite{wax}(which again poses a difficult astrophysical problem) and the fact that in Z decay there are many more pions
(hence photons from
$\pi^0$ decay) than protons. Another class of proposed explanations, which I find more appealing, is to invoke the
decay of some superheavy particle of mass $M\gappeq 10^{12}~GeV$. It could also be a topological defect \cite{ber}. But
a particle candidate would be a cosmion \cite{sar}, an almost completely stable particle (lifetime longer than the
universe life) with only gravitation interactions, possibly from a hidden sector, a remnant of the quantum gravity
world, with relatively small mass in comparison to $M_{GUT}$ in order for its density not to be diluted by inflation.
This particle would contribute to the dark matter expecially clustered near galaxies like ours. Its rare decays would
generate the observed protons. Again why so many protons and not even more pions? Advocates of these solution argue
that we  only have experience with the final state of objects of mass of order 100 GeV, not
$10^{12}~GeV$ or more.

\section{Conclusion}

Today in particle physics we follow a double approach: from above and from below. From above there are, on the theory
side, quantum gravity (that is superstrings), GUTs and cosmological scenarios. On the experimental side there
are underground experiments (e.g. searches for neutrino oscillations and proton decay), cosmic ray
observations, satellite experiments (like COBE, IRAS etc) and so on. From below, the main objectives of theory and
experiment are the search of the Higgs and of signals of particles beyond the Standard Model (typically supersymmetric
particles). Another important direction of research is aimed at the exploration of the flavour problem: study of CP
violation and rare decays. The general expectation is that new physics is close by and that should be found very
soon if not for the complexity of the necessary experimental technology that makes the involved time scale painfully
long. 

I am very grateful to Professor Oscar Saavedra for his kind invitation and hospitality.

\end{document}